# Enhanced superconducting performance of melt quenched $Bi_2Sr_2CaCu_2O_8$ (Bi-2212) superconductor

Jagdish Kumar[1,2], Devina Sharma[1,3], P.K. Ahluwalia[2] and V.P.S. Awana[1,*]

[1]*Quantum Phenomena and Application Division, National Physical Laboratory, New Delhi-110012, India*

[2]*Department of Physics, Himachal Pradesh University, Shimla-171005, India*

[3]*Department of Physics, Punjab University, Chandigarh, India – 160014*

## ABSTRACT

We scrutinize the enhanced superconducting performance of melt quench Bismuth based $Bi_2Sr_2CaCu_2O_8$ (Bi-2212) superconductor. The superconducting properties of melt quenched Bi-2212 (Bi2212-MQ) sample are compared with non-melted Bi2212-NM and $Bi_{1.4}Pb_{0.6}Sr_2Ca_2Cu_3O_{10}$ (Bi-2223). Crystal structure and morphology of the samples are studied using X-ray diffraction and Scanning Electron Microscopy (SEM) techniques. The high field (14T) magneto-transport and DC/AC magnetic susceptibility techniques are extensively used to study the superconducting properties of the investigated samples. The superconducting critical temperature ($T_c$) and upper critical field ($H_{c2}$) as well as thermally activated flux flow (TAFF) activation energy are estimated from the magneto-resistive [R(T)H] measurements. Both DC magnetization and amplitude dependent AC susceptibility measurements are used to determine the field and temperature dependence of critical current density ($J_c$) for studied samples. On the other hand, the frequency dependent AC susceptibility is used for estimating flux creep activation energy. It is found that melt quenching significantly enhances the superconducting properties of granular Bi-2212 superconductor. The results are interpreted in terms of better alignment and inter-connectivity of the grains along with reduction of grain boundaries for Bi2212-MQ sample.

*: Corresponding author: awana@mail.nplindia.ernet.in;
Web page: www.freewebs.com/vpsawana

# INTRODUCTION

As far as fabrication of superconducting wire and tapes is concerned, despite the growing popularity of high temperature superconducting thin films and single crystals, the polycrystalline bulk materials still holds importance [1] [2] [3] [4]. The high temperature cuprate superconductors (HTSc) offer highest superconducting critical temperatures values among all known classes of superconducting materials [5] [6] [7] [8]. This makes them most desired materials for various applications of superconducting properties at higher temperatures and even at higher applied magnetic fields [9]. However, one major hindrance to realize them in industrial applications is their granular nature and small coherence length [10]. Due to the small value of their coherence length, the grain boundaries in these materials act as sort of weak links and thus limit the superconducting performance of these materials in polycrystalline form [11][12]. Moreover, these materials are strongly anisotropic and the value of coherence length in *ab* plane is significantly larger than that along *c*-direction [13] [14] [15]. Due to this fact the polycrystalline samples having aligned grains along *c*-axis have significantly better inter-granular coupling and thus superior superconducting performance [16] [17] [18]. The melt quench technique offers a good means to improve inter-grain connectivity due to reduced porosity and possible crystallographic alignment of grains [16][18][19].

Bismuth based cuprate superconductors are among promising candidates for wires and tapes applications due to higher values of their critical temperature, current density and critical magnetic field [13][14][20][21][22]. Three main phases of BSCCO superconductor are known as Bi-2201($Bi_2Sr_2CuO_6$), Bi-2212($Bi_2Sr_2CaCu_2O_8$) and Bi-2223 ($Bi_{1.4}Pb_{0.6}Sr_2Ca_2Cu_3O_{10}$) having $T_c$ values of around 15, 85 and 110K respectively [23]. Among these phases Bi-2201 has lower $T_c$ and thus of limited interest to applications in tapes and wires. Bi-2223 has highest value of critical temperature, but is difficult to obtain in single phase and generally results with intergrowth of Bi-2212 phase and also requires long hours (>100 hrs) sintering for phase growth [24][25]. On the other hand Bi-2212 is relatively easy to synthesize and hence can be obtained in majority single phase form. Thus we have here studied effect of melt quenching on superconducting performance of Bi-2212 phase and compared its superconducting performance with non melted (Bi2212-NM) and Bi-2223. Extensive magneto-resistive and DC/AC magnetic susceptibility measurements are carried on various samples (Bi2212-NM, Bi2212-MQ, Bi-2223) to study and compare various superconducting properties like flux pinning, temperature and field dependence of critical current density, upper critical field, flux creep and thermally activated flux flow activation energy. It is found that superconducting performance of Bi-2212 enhances significantly by melt quenching and it can be even better than a polycrystalline Bi-2223 sample.

**EXPERIMENTAL DETAILS**

The samples of nominal composition $Bi_2Sr_2CaCu_2O_8$ (Bi-2212) and $Bi_{1.4}Pb_{0.6}Sr_2Ca_2Cu_3O_{10}$ (Bi-2223) were synthesized by solid state reaction route starting from high purity powders of $Bi_2O_3$, PbO, $SrCO_3$, $CaCO_3$ and CuO. The reactants were weighed in stoichiometric proportions and were ground properly to obtain homogeneously mixed powders. The obtained mixture was calcined at 800°C for 12 hours and reground after cooling. The calcined samples were then annealed in air at 820, 840 and 860°C for 12 hours with intermediate regrinding. Finally, the obtained powders were pressed into rectangular bar shaped pallets. The pallets of Bi-2223 were sintered at 850°C for 240 hours. The pallets of Bi-2212 were divided into two sets. One set of the pallets were melted partially at 930°C for 8-10 minutes and then quenched to room temperature. These samples were termed as Bi2212-MQ and will be called MQ hereafter. The second set was termed as Bi2212-NM and will be called NM hereafter. Both NM and MQ samples were finally sintered at 860°C for 24 hours.

The obtained samples were then examined for their structural properties by measuring their room temperature X-ray diffraction pattern using Rigaku Miniflex X-ray diffractometer with Cu-K$_\alpha$ radiation. The micro-structural studies were performed using scanning electron microscopy (SEM). Magneto-transport measurements were carried out using standard four probe technique using Quantum Design Physical Property Measurement System (PPMS) in the field range of 0-14T. The temperature dependent *DC* and *AC* susceptibility measurements were done using ACMS facility being employed in PPMS. These measurements were carried out for the samples in the temperature range of 10 to 120K with amplitude and frequency of applied field varying from 0.5 to 11Oe and 33 to 9999Hz respectively.

## RESULTS AND DISCUSSIONS
### Structural and morphology studies

Figure 1 (a) represents the room temperature X-ray diffraction (XRD) pattern of the Bi$_2$Sr$_2$CaCu$_2$O$_8$ non-melted (NM) and melt quenched (MQ) samples. It can be seen that all the characteristic peaks corresponding to Bi-2212 phase are found in the XRD pattern [26]. However, in addition some peaks of Bi-2201 phase are also detected in Bi2212-MQ sample. This shows that the Bi-2212 starts to degrade to lower $T_c$ Bi-2201 phase if subjected to higher temperature for longer time. Thus it is important to perform melt quenching carefully so as to optimize inter-granular connectivity and reduce degradation of Bi-2212 phase to Bi-2201. We did not find any significant change in peak positions along 2θ axis for NM or MQ samples, which show that lattice parameters are almost same for both the samples (see Table 1). Thus any sort of structural distortion during melt quenching can be ruled out. Moreover, it is seen that peaks corresponding to planes along [001] direction, are of higher intensity for MQ sample as compared to NM sample. This indicates preferred alignment of grains along *c*-axis for MQ sample in comparison to NM sample [27].

Figure 1b shows the XRD pattern for Bi-2223 sample. It is found that in addition to characteristic peaks corresponding to Bi-2223 phase, the peaks of Bi-2212 and $CaPb_2O_4$ phase are also observed [28]. Usually Bi-2223 phase is obtained with significant amount of Bi-2212 phase [24] [25] [28]. This is generally expected in the BSCCO superconductors, since intergrowth of its various phases viz. 2201, 2212 and 2223 are observed. However, in the present case the Bi-2223 peaks are relatively intense showing larger fraction of this phase in the sample. The phase percentage of Bi-2223 in mixture was determined using equation [29]:

$$X_{Bi2223} = \frac{I_{Bi2223}^{0\ 0\ 10}}{0.88 I_{Bi2212}^{0\ 0\ 8} + I_{Bi2223}^{0\ 0\ 10}} \quad \ldots\ldots (1)$$

and is found to be around 63 percent.

Figure 2 shows SEM images of Bi2212-NM, Bi2212-MQ and Bi-2223 samples at 5kX magnification. The granular nature of Bi2212-NM and Bi-2223 samples is clearly visible in the SEM images. It is noteworthy that grains of latter seem to be fused and larger than those from former. On the other hand, it can be seen that Bi2212-MQ is almost a dense melt. In Bi-based superconductors, the double bismuth oxide layers are connected by weak Vander Waal's forces due to which the grains can be easily aligned during process of melt quenching. Thus melt quenching results in highly oriented well connected grains with significantly reduced number of grain boundaries. This fact is evidenced from our XRD results that show *c*-axis alignment of the grains in melt quenched samples.

**Magneto-transport studies**

The resistance of the studied samples normalised to room temperature is shown in Figure 3. It can be seen that all the samples are metallic up to $T_c^{onset}$ below which they exhibit superconducting behaviour. The superconducting onset was found to be same i.e. ~93K for both MQ and NM while ~116K for Bi-2223 sample. The small drop in resistance for Bi2212-NM sample near 116K can be attributed to presence of small content of Bi-2223 phase that may not be detectable by XRD measurements. It is to be noted that resistive drop corresponding to Bi-2223 is not observed for MQ sample. This may be because of degradation of Bi-2223 to lower $T_c$ phases during the melt quenching. The inset of Figure 3 shows the magnified view of transition near zero resistance temperature $T_c^{R=0}$. The values of $T_c^{R=0}$ and superconducting transition width, $\Delta T = T_c^{onset} - T_c^{R=0}$ for NM, MQ and Bi-2223 samples are given in Table 2. It is interesting to note that the superconducting transition for MQ sample is sharpest among all the studied samples. Also, it can be seen from the temperature dependence of normal state resistance that MQ and Bi-2223 samples are more metallic in comparison to NM sample. Thus better inter-grain connectivity in former can be expected than in latter. Thus it is evident that melt quenching facilitates formation of highly aligned gains along with better inter-grain connectivity.

Figures 4 shows temperature dependence of resistance for studied samples in applied magnetic fields ranging from 0 to 14T. It is clearly visible that while the onset of superconducting transition remains almost same for all the samples, the $T_c^{R=0}$ shifts by large value. The shift, $T_c^{R=0}(0T)-T_c^{R=0}(14T)$ in zero resistive transition temperature with applied field is found to be 54K, 46K and 61K for NM, MQ and Bi-2223 samples respectively.

Since, the presence of applied magnetic field tends to weaken the connectivity between the grains and even decouple them at sufficiently high field values, transition near zero resistivity spreads with applied field. Hence, the least shift of 46K in $T_c^{R=0}$ for MQ sample shows that the inter-grain coupling is best in it among all the studied samples.

The magneto-resistive curves are used to calculate the temperature dependence of upper critical field ($\mu_oH_{c2}$(T)) for all the samples. $H_{c2}$ is defined as the field for which the temperature dependent resistance R($H_{c2}$,T)=0.9R$_n$, where R$_n$ is the normal state resistance. Conventional single band WHH theory [30] which describes the orbital limited upper critical field of dirty type II superconductors is used to fit the obtained data using equations,

$$\ln(1/t) = \psi\left(\frac{1}{2} + \frac{\bar{h}}{2t}\right) - \psi(1/2) \quad \ldots\ldots\ldots (2)$$

where, $t=T/T_c$ is reduced temperature, $\Psi$ is a digamma function and $\bar{h}$ is given by

$$\bar{h} = \frac{4H_{c2}}{\pi^2 T_c}\left(\frac{-dH_{c2}}{dT}\right)_{T=T_c} \quad \ldots\ldots\ldots (3)$$

Figure 5 shows the $\mu_oH_{c2}(T)$ of the three samples determined from 90% of $\rho_n(T)$ in inset and its fitting using simplified WWH formula. The low temperature values of $H_{c2}(T)$ are obtained by extrapolating WWH formula using equations (2) and (3). Thus, values of $\mu_oH_{c2}(0)$ are estimated to be 104.7, 518.9 and 411.9T for NM, MQ and Bi-2223 samples respectively. It is interesting to notice that upper critical field for MQ sample is highest. We speculate this enhancement in value of $H_{c2}(0)$ of MQ sample is due to c-axis alignment of its grains, which is evident from our XRD and SEM results. It is well known that BSCCO systems are highly anisotropic with in-plane and out of plane coherence length values of ~0.9 and 0.05nm respectively [13]. Thus alignment of grains along c-axis will result in decrease of average coherence length of polycrystalline MQ sample, when field is applied along axis of preferred orientation, thereby, resulting in enhancement of its $H_{c2}(0)$. The zero temperature coherence length $\xi(0)$ is estimated from $\mu_oH_{c2}(0)$ using Ginzburg-Landau formula given by $\mu_oH_{c2}(0)=\varphi_o/[2\pi\,\xi^2(0)]$, where, $\varphi_o$ is given by $2.07\times10^{-15}$Wb. All the obtained parameters are listed in Table 1.

The broadening of the resistive transitions (as seen in figure 4), in the presence of magnetic fields near $T_c^{R=0}$ are interpreted in terms of energy dissipation caused vortex motion. The flow of vortices in this region is caused by thermal activation and hence, is known as thermally activated flux flow (TAFF) [12]. Thus, resistivity in this region is given by well known Arrhenius equation, $\rho = \rho_0(B,T)e^{-U_0/\kappa_BT}$ [31][32], where $U_0$ is the TAFF activation energy, which can be obtained from the slope of the linear part of the ln ($\rho/\rho_0$) versus $T^{-1}$ plot. Here, $\rho_0$ has been taken as normal state resistance at 100K for both NM and MQ and 120K for Bi-2223 samples. Figure 6 shows the ln ($\rho/\rho_0$) versus $T^{-1}$ plot for the studied samples. $U_0$ is obtained from the limited range of data yielding straight line in the low resistivity region. The corresponding linear scale fitting done to obtain $U_0$ at various fields is shown in the same figure. Figure 7 shows the obtained activation energy of the studied samples at various fields in log-log scale. It is found that TAFF activation energy scales as power law ($U_0=K\times H^{-\alpha}$) with magnetic field [32]. Also, it can be seen that field dependence of $U_0$ is different for lower and higher field values. The values of parameters obtained from power law fitting at lower ($K_{LF}$, $\alpha_{LF}$) and higher magnetic field ($K_{HF}$, $\alpha_{HF}$) as well as $U_0$ at 0K for the studied samples are shown in table 2. Similar dependence of exponents in power law with field has been reported by Palstra et al. for Bi-2212 single crystals [32][33]. Interestingly for both Bi2212-NM and Bi2212-MQ samples there is a change in activation energy behaviour, while going from low to high field region but no significant change for Bi-2223 sample. Moreover, activation energy of flux creep is higher for MQ than the NM sample, evidencing the stronger flux pinning for the former than later.

# Magnetic susceptibility studies

## (a) DC susceptibility

The DC magnetisation versus applied magnetic field (*M-H*) measured at 5K is shown in Figure 8 for all the studied samples. From *M-H* data we have calculated the value of critical current density using Bean's critical state model [34]. According to this model the critical current density for an infinitely long sample having rectangular cross section (a × b) (with a < b) is given by $J_c(H)=[20\times\Delta M(H)]/[a(1-a/3b)]$, where ΔM is the width ΔM of the M(H) hysteresis loop [10] [35]. Figure 9 shows the $J_c(H)$ curves obtained for all the samples using above formula. The $J_c$ value is highest (~34kA/cm$^2$ for low magnetic field) for Bi-2223 sample throughout the measured field. Also it is interesting to note the large difference in $J_c$ of MQ (~16kA/cm$^2$) sample in comparison to that of NM (~4kA/cm$^2$) sample of Bi-2212. The considerably improved value of $J_c$ of MQ sample can be attributed to the alignment of the grains and reduction of number of grain boundaries as evidenced from XRD and SEM analysis. Systematic variation of microstructure has been carried out recently by various groups [11] [12] [36] who have found that superconducting properties are affected significantly with microstructure of the samples. Thus it is worth mentioning that microstructure plays crucial role in superconducting properties of BSCCO superconductors.

One of the characteristics of type II superconductors is the existence of vortices when applied field is increased above lower critical field $H_{c1}$. These vortices tend to move under applied magnetic field due to Lorentz force acting on them, resulting in Ohmic losses [10]. Thus to achieve practically lossless currents, these vortex lines must be pinned to avoid flux line motion. The average pinning force density that balances the Lorentz force acting on vortices is given by [10] $F_P = J_c \times B$, where $J_c$ is critical current density. The pinning force density calculated from $J_c(H)$ data is given in Figure 10. Clearly, the pinning force density $F_p$, is maximum for Bi-2223 sample. On comparing NM and MQ samples (see Table 1), we find that $F_P$ for MQ is significantly higher than for NM.

**(b) AC susceptibility**

Figure 11 (a) and (b) shows amplitude dependence (0.5 to 11.0Oe) of real ($\chi'$) and imaginary part ($\chi''$) of AC susceptibility (ACS) respectively, studied as a function of temperature, for all the mentioned samples. For the weakly coupled polycrystalline superconductors, as temperature is lowered, real part of susceptibility ($\chi'$) shows two transitions corresponding to intra and inter grain coupling, accompanied by corresponding loss peaks in imaginary part ($\chi''$) [10]. Therefore, comparing we find from Figure 11 (a) that why both NM and Bi-2223 samples shows signatures of granularity by exhibiting two step transition while MQ sample shows single step transition indicating good grain connectivity. Also in case of polycrystalline samples total susceptibility results from contribution of the superconducting grains as well as non superconducting matrix [37]. Hence saturation value of $\chi'$ represents total superconducting volume fraction of the sample. Using this criterion superconducting volume fraction of NM, MQ and Bi-2223 sample comes out to be 70%, 73%, 90% respectively at 0.5Oe. However, it is noteworthy that with an increase of applied magnetic field amplitude, the saturated susceptibility reduces considerably for NM and Bi-2223 samples, while the change is minute for MQ sample. This indicates best inter-grain connectivity for MQ sample at high field magnitudes.

It can be seen from Figure 11 (b) that the loss peak observed in $\chi''$, shifts to lower value of temperature when amplitude of applied field increases. As described above, this loss peak corresponds to energy losses in the sample per cycle of the drive signal, which can be attributed to response of vortices to the external field. In the transition region, the sample is not fully superconducting and thus field can easily penetrate and has to get out of the sample with field reversal during each cycle in ACS. This results in loss peak observed in $\chi''$ at temperatures corresponding to intra/inter granular transition. The peak in $\chi''$ is observed when field penetrate up to centre of the sample and thus maximum loss occurs. The separation of intra and inter-granular loss peak on temperature scale indicates weakly coupled grains and can be seen for NM and Bi-2223 samples. Whereas for MQ sample both intra and inter-granular loss peak merges, showing strong coupling between the grains. The shifting of the loss peaks with applied field amplitude is used to calculate the temperature dependence of critical current density ($J_c$). According to Bean's critical state model, the $J_c$ at peak temperature $T_p$ for a bar shaped sample having cross action 2a×2b can be determined by [38]:

$$J_c(T_p) = H^*/\sqrt{ab} \quad \ldots\ldots (4)$$

where, $H^*$ is the amplitude of applied magnetic field. Inter granular current density obtained using above equation is given in Figures 12 (a) to (c). The temperature dependence of $J_c$ [39][40] is found to obey power law given by

$$J_c(T) = J_c(0)(1-T_p/T_c)^n \quad \ldots\ldots (5)$$

where, $T_c$ is the zero resistivity critical temperature. Values of $J_c(0)$ and '$n$' obtained from fit to experimental data are provided in Table 1. It is interesting to note that $J_c(0)$ is highest for MQ sample and is almost three orders of magnitude higher than NM or Bi-2223 samples, which is in contrast to *M-H* measurements given in Figure 9. The highest value of $J_c$ obtained from low fields ACS measurements can be explained on the basis of Figure 7, which shows a crossover of activation energy of MQ and Bi-2223 samples at around 500Oe. At lower field values, MQ has highest activation energy, while it is highest for Bi-2223 at higher fields. This indicates strongest pinning for MQ samples at low fields and at higher fields for Bi-2223 sample. Moreover, the value of *n* is highest for MQ sample, indicating fastest decrease of $J_c$ with temperature as compared to NM or Bi-2223 samples. Hence, it strengthens our speculation that there is crossover of $J_c$ for MQ and Bi-2223 samples from lower to higher field values respectively. Table 1 shows the obtained values of $J_c(0)$ and $n$ for the studied samples.

Figure 13 (a) and (b) shows frequency dependence (33 to 9999Hz) of real ($\chi'$) and imaginary ($\chi''$) parts of AC susceptibility studied as a function of temperature. The dissipative peak in $\chi''$ shifts towards higher value of temperature as the frequency is increased. The peak in $\chi''$ corresponds to energy loss when the applied field penetrates till the centre of the sample. With the increasing frequency, magnetic field has less time to penetrate the sample and thus sample must have weaker pinning in order to reach full penetration and hence peak in $\chi''$. Since, pinning force is inversely proportional to temperature therefore, loss peak in $\chi''$ shifts to higher temperature. From the shift in $\chi''$ peak temperature ($T_p$) with frequency ($f$) of AC field, one can calculate activation energy $E_a$ required for flux creep [41] by using equation given by Nikolo and Goldfarb [42]:

$$f = f_0 exp\left[\frac{-E_a}{K_B T_P}\right] \quad \ldots\ldots (6)$$

where, $f_0$ is some constant with dimensions of frequency and $k_B$ is Boltzmann constant. This equation can also be written as:

$$\ln(f) = \ln f_0 - \frac{E_a}{K_B T_P} \quad \ldots\ldots (7)$$

Thus by plotting $\ln(f)$ versus $1/T_P$ one should obtain straight line, whose slope when multiplied by Boltzmann constant gives activation energy. Figure 14 (a) to (c) gives the plots between $\ln(f)$ and $1/T_P$ and corresponding data for linear fitting for the studied samples. From these fittings the activation energy was found to be 59, 737, and 520 meV for NM, MQ and Bi-2223 respectively. It is worth mentioning that the value of activation energy for MQ is largest. This indicates strongest vortices pinning in MQ sample and thus enhanced superconducting performance as has been observed.

## CONCLUSIONS

In conclusion, we have studied three different samples of BSCCO superconductors, viz. Bi2212-NM, Bi2212-MQ and Bi-2223. It is found that though long span melting of Bi-2212 sample results in its degradation to lower $T_c$ Bi-2201 phase, the small duration partial melting results in good inter-grain connectivity and crystallographically aligned grains. This results in relatively sharper resistive transition and enhancement of superconducting properties in case of Bi2212-MQ sample. The characteristic superconducting parameters like flux pinning, thermally activated flux flow, flux creep activation energies, and the $J_c$ and $H_{c2}$ are improved significantly for Bi2212-MQ sample. It is thus envisaged that partially melted Bi2212-MQ could be a good candidate for bulk superconducting applications of high $T_c$ cuprates.

## ACKNOWLEDGEMENTS

JK would like to acknowledge CSIR for providing financial assistance in the form of SRF. Authors thank Prof. R. C. Budhani, Director, NPL for his keen interest and encouragement in superconductivity research. We also acknowledge Mr. A. K. Sood from SEM division for providing the SEM images of the samples.

**Figure Captions**

**Figure 1** XRD patterns of (a) Bi2212-NM and Bi2212-MQ and (b) Bi-2223 samples.

**Figure 2** SEM images of (a) Bi2212-NM, (b) Bi2212-MQ and (c) Bi-2223 samples taken at 5kX magnification.

**Figure 3** Temperature dependence of resistance of the studies samples normalized to room temperature value. The inset shows values close to zero resistance regions.

**Figure 4** Temperature dependence of normalized resistance for applied magnetic field ranges 0-14T.

**Figure 5** Temperature dependence of upper critical field (symbols) and fitting (solid line) done using simplified WHH theory. Inset shows the corresponding experimental $\mu_o H_{c2}(T)$ derived from magneto-resistive transitions.

**Figure 6** The Arrhenius plots of electrical resistance for fields from 0-14T. The red lines represent linear fitted data in low resistance regions.

**Figure 7** Magnetic field dependence of TAFF activation energy plotted on log-log scale. The lines represent power law fitting to the experimental data.

**Figure 8** Field dependence of *DC* magnetization at 5K temperature.

**Figure 9** Field dependence of critical current density for studied samples measured at 5K temperature.

**Figure 10** Magnetic field dependence of pinning force density of studied samples.

**Figure 11(a)** Temperature dependence of real part of AC susceptibility studied for various amplitudes of applied field.

**Figure 11(b)** Imaginary part of AC susceptibility as function of temperature for studied samples at various amplitudes of applied field.

**Figure 12** Temperature dependence of critical current density (symbols) and its fitted curve

(solid line) as per equation, (5) for (a) Bi2212-NM, (b) Bi2212-MQ and (c) Bi-2223 sample.

**Figure 13** Temperature dependence of (a) real and (b) imaginary part of AC susceptibility for studied samples at different frequencies of applied AC field.

**Figure 14** Linear fitting curves (solid line) of $ln(f)$ versus $1/T_p$ data (symbols) for (a) Bi2212-NM, (b) Bi2212-MQ and (c) Bi-2223 samples. The flux creep activation energy is obtained from slope of these fitted lines.

**Table 1:** Superconducting parameters obtained from DC and AC susceptibility measurements

| Sample | $a$ (Å) | $c$ (Å) | $J_c$ @5K (kA/cm$^2$) | $F_P$ @5T (N/m$^3$) | $E_a^{ACS}$ @3.0Oe (meV) | $J_c$ @0K kA/cm$^2$ | $n$ |
|---|---|---|---|---|---|---|---|
| Bi2212-NM | 3.83 | 30.95 | 4 | 5×10$^7$ | 59 | 0.154 | 3.85 |
| Bi2212-MQ | 3.83 | 30.83 | 16 | 2.4×10$^8$ | 737 | 596.061 | 4.85 |
| Bi-2223 | 3.82 | 37.12 | 34 | 4.7×10$^8$ | 520 | 0.656 | 3.52 |

**Table 2:** Superconducting parameters obtained from magneto-resistive measurements

| Sample | $T_c^{onset}$ (K) | $T_c^{R=0}$ (K) | $\Delta T_c$ (K) | $\mu_0 H_{c2}$ @0K (T) | $\xi(0)$ (nm) | $K_{LF}$ | $\alpha_{LF}$ | $K_{HF}$ | $\alpha_{HF}$ | $E_a^{RTH}$ @0Oe (meV) |
|---|---|---|---|---|---|---|---|---|---|---|
| Bi2212-NM | 93 | 80 | 13 | 104.7 | 1.7 | 568 | 0.33 | 559 | 0.15 | 286 |
| Bi2212-MQ | 93 | 85 | 8 | 518.9 | 0.8 | 560 | 0.86 | 1473 | 0.42 | 3557 |
| Bi-2223 | 116 | 103 | 13 | 411.9 | 0.9 | 2321 | 0.33 | 2066 | 0.35 | 1555 |

# REFERENCES

[1] J. L. H. Lindenhovius, E. M. Hornsveld, A. Den Ouden, W. A. J. Wessel, H. H. J. Ten Kate, IEEE Transactions on Applied Superconductivity **10** (2000) 975

[2] B. A. Glowacki, M. Majoros, M. E. Vickers, B. Zeimetz, arXiv:cond-mat/0109085v1 (2001)

[3] T. P. Beales, J. Jutson, L. L. Lay and M. Molgg, J. Mater. Chem. **7** (1997) 653

[4] Y. Ma, Z. Gao, Y. Qi, X. Zhang, L. Wang, Z. Zhang, Dongliang Wang, Physica C **469** (2009) 651

[5] P. Dai, B.C. Chakoumakos, G.F. Sun, K.W. Wong, Y. Xin, D.F. Lu, Physica C **243** (1995) 201

[6] Shilling A., Cautoni M., Gao J. D., Ott H. R., Nature **363** (1993) 56

[7] Chu C. W., Gao L., Chen F., et al., Nature **365** (1993) 323

[8] Daminov R. R., Imayev M. F., Reissner M., et al., Physica C **46** (2004) 408

[9] A. Ballarino, L. Martini, S. Mathot, T. Taylor, and R. Brambilla, IEEE transactions on applied superconductivity **17** (2007) 3121

[10] G. Krabbes, G. Fuchs, W. Canders, H. May, R. Palka, "*High Temperature Superconductor Bulk Materials*", Wiley-VCH, Weinheim 9 (2006)

[11] Devina Sharma, Ranjan Kumar, H. Kishan and V.P.S. Awana, J Supercond Nov Magn, **24** (2011) 205

[12] Devina Sharma, RanjanKumar and V.P.S.Awana, Solid State Communications **152** (2012) 941

[13] A. Pomar, M. V. Ramallo, J. Mosqueira, C. Torron, and Felix Vidal, Phys. Rev B **54**, (1996) 7470

[14] S. I. Vedeneev, A. G. M. Jansen, E. Haanappel, P. Wyder Phys. Rev. B **60** (1999) 12467

[15] Melissa Charalambous, Jacques Chaussy, and Pascal Lejay, Phys. Rev. B **45** (1992) 5091

[16] A. Matsumoto, H. Kitaguchi, H. Kumakura, and K. Togano, IEEE transactions on applied superconductivity **11** (2001) 3046

[17] S. Li, Q.Y. Hu, H.K. Liu, S.X. Dou and W. Gao, Physica C **279** (1997) 265

[18] H. Kumakura, H. Kitaguchi and K. Togano, T. Muroga and J. Sato and M. Okada, IEEE transactions on applied superconductivity, **9** (1999) 1804

[19] V P S Awana, S B Samanta, P K Dutta, E Gmelin and A V Narlikar, J. Phys.: Condens. Matter **3** (1991) 8893

[20] Ken-ichiro Takahashi, Takayuki Nakane, Akiyoshi Matsumoto, Hitoshi Kitaguchi, and Hiroaki Kumakura, IEEE transactions on applied superconductivity, **19** (2009) 3067

[21] Jean-Michel Rey, Arnaud Allais, Jean-Luc Duchateau, Philippe Fazilleau, Jean-Marc Gheller, Ronan Le Bouter, Olivier Louchard, Lionel Quettier, and Daniel Tordera, IEEE transactions on applied superconductivity, **19** (2009) 3088

[22] T. Kagiyama et al. Materials Science and Engineering **18** 152001 (2011)

[23] J. L. Tallon, R. G. Buckley, P. W. Gilberd, M. R. Presland, I. W. M. Brown, M. E. Bowden, L. A. Christian & R. Goguel, Nature **333** (1988) 153

[24] E. Chavira, R. Escudero, D. Rios Jara and L. M. Leon Phys. Rev. B **38** (1988) 9272

[25] A. Polasek, P. Majewski, E. T. Serra, F. Rizzo and F. Aldinger, Materials Research **7** (2004) 393

[26] J. Kumar, P.K. Ahluwalia, H. Kishan and VPS Awana, J. Supercond. Nov. Magn. **23** (2010) 493

[27] G. Yildirim S. Bal and A. Varilci, J. Supercond. Nov. Magn. doi:10.1007/s10948-012-1497-1 (2012)

[28] Huseyin Sozeri, Nader Ghazanfari, Husnu Ozkan and Ahmet Kilic, Supercond. Sci. Technol. **20** (2007) 522

[29] Q.Y. Hu, H.K. Liu, S.X. Dou, Physica C **250** (1995) 7

[30] N. R. Werthamer K. Helfand and P. C. Hohenberg, Phys. Rev. **147** (1966) 295

[31] Dew Hughes, Cryogenics **28** (1988) 674

[32] T. T. M. Palstra, B. Batlogg, R. B. van Dover, I. F. Schneemeyer and J. V. Waszczak, Phys. Rev. B **41** (1990) 6621

[33] T. T. M. Palstra, B. Batlogg, L. F. Schneemeyer, and J. V. Waszczak, Phys. Rev. Lett. **61** (1988) 1662

[34] C. P. Bean, Phys. Rev. Lett. **8** (1962) 250

[35] C S Yadav and P L Paulose, New J. Phys. **11** (2009) 103046

[36] A. A. Khurram, M. Mumtaz, Nawazish A. Khan, M. M. Ahadian and Azam Iraji-zad, Supercond. Sci. Technol. **20** (2007) 742

[37] F Gomory, Supercond. Sci. Technol. **10** (1997) 523

[38] H. Salamati, P. Kameli, Physica C **403** (2004) 60

[39] Vinay Ambegaokar and Alexis Baratoff, Phys. Rev. Lett. **10** (1963) 486

[40] P. O. de Gennes, Rev. Mod. Phys. **36** (1964) 225

[41] Mortady I. Youssif, Egypt. J. Sol., Vol. **25** (2002) 171

[42] Nikolo M. And Goldfrab R. B., Phys. Rev. B **39** (1989) 6615

**Figures:**

Figure 1 (a) and (b)

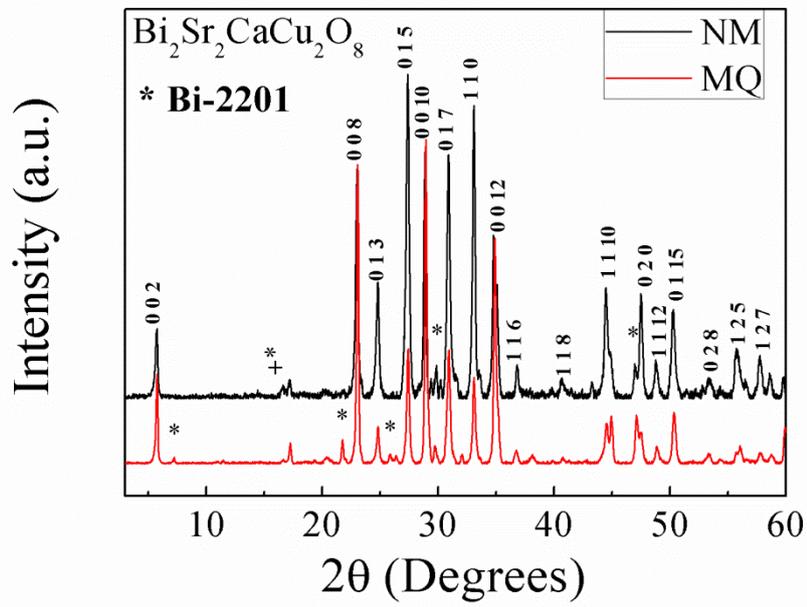

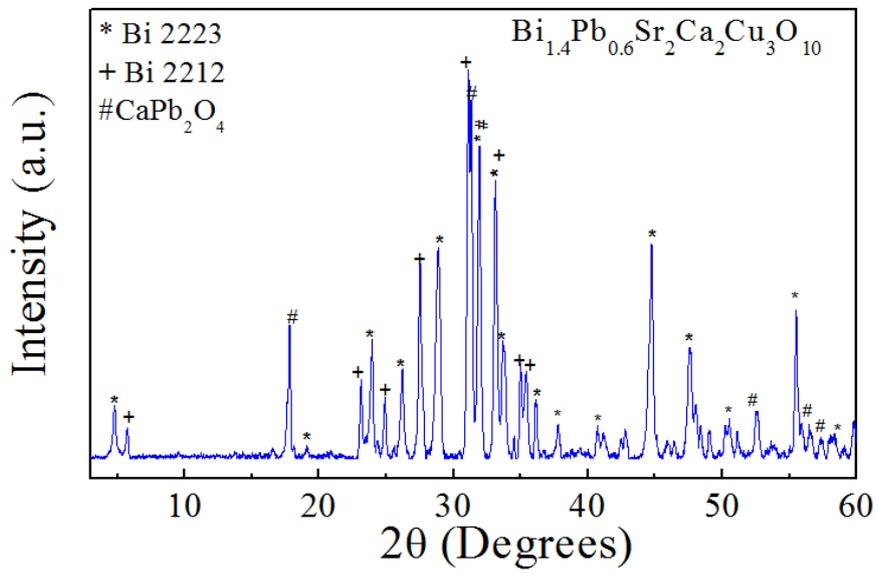

Figure 2(a) to (c)

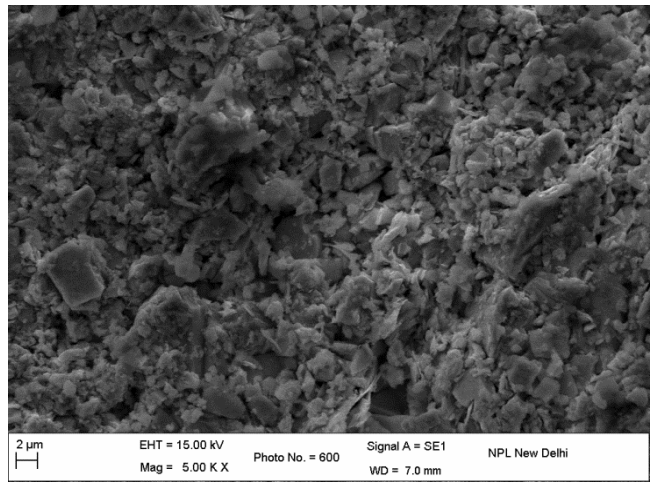

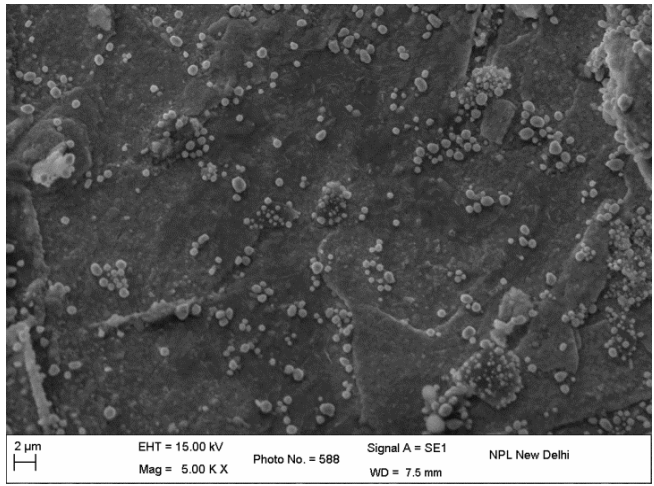

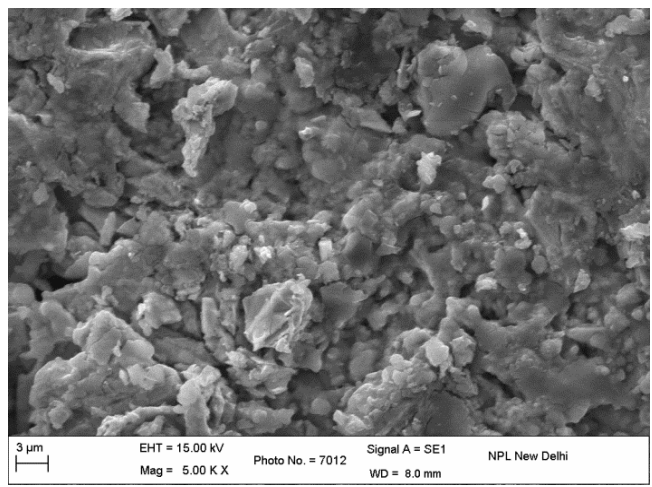

Figure 3

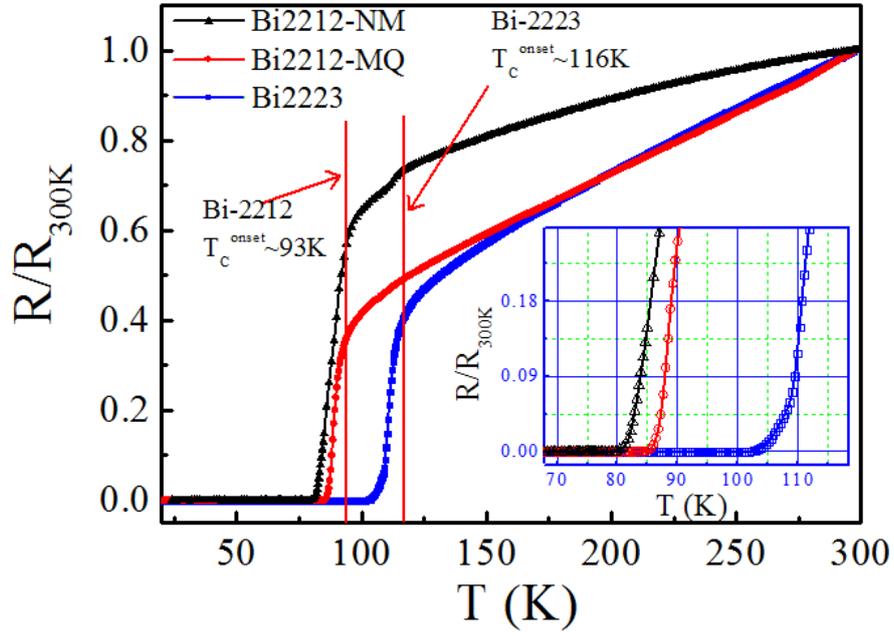

Figure 4

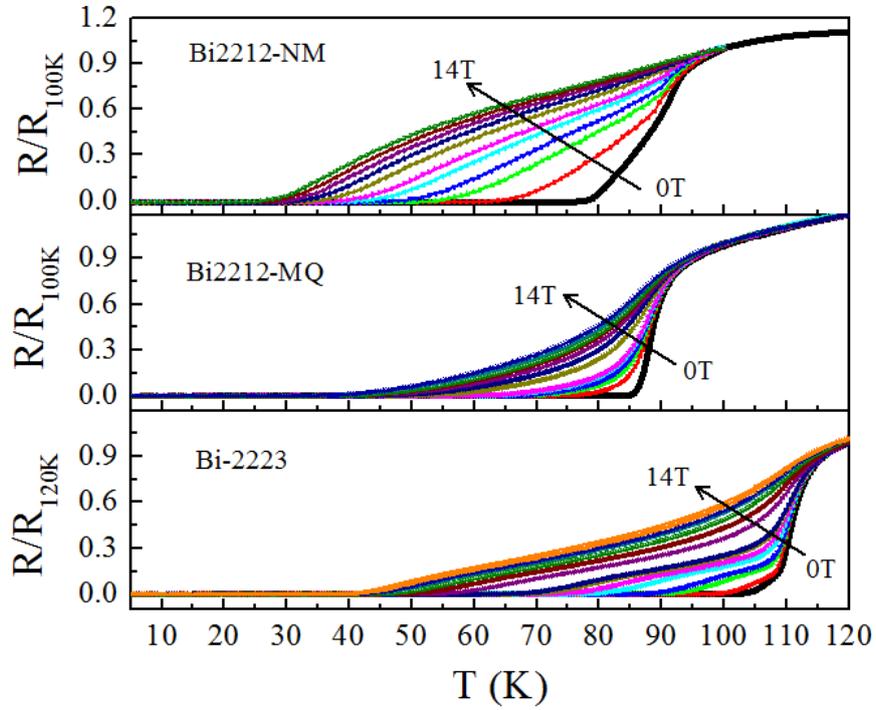

Figure 5

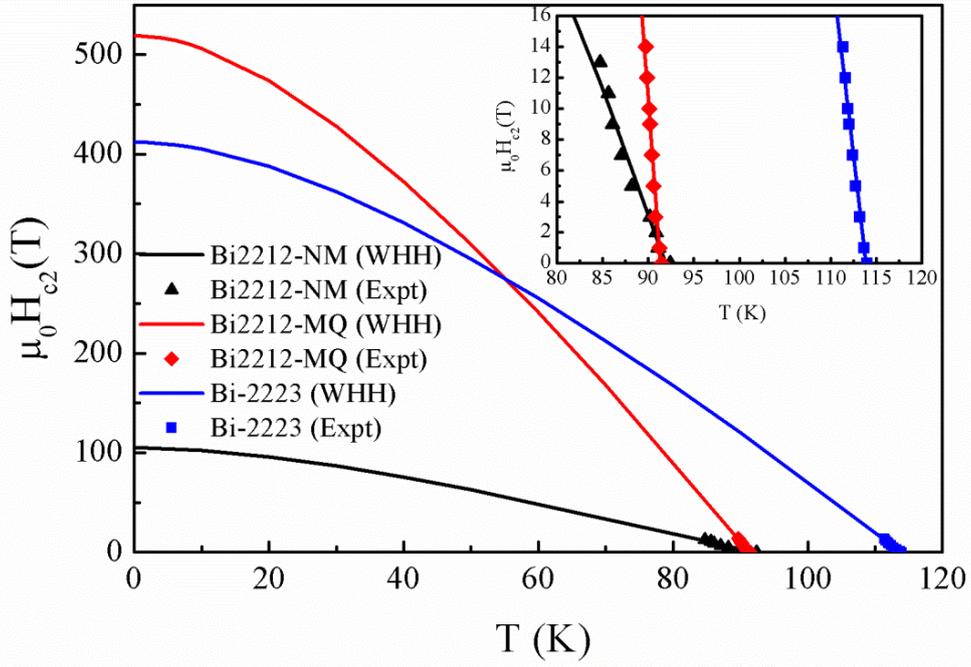

Figure 6

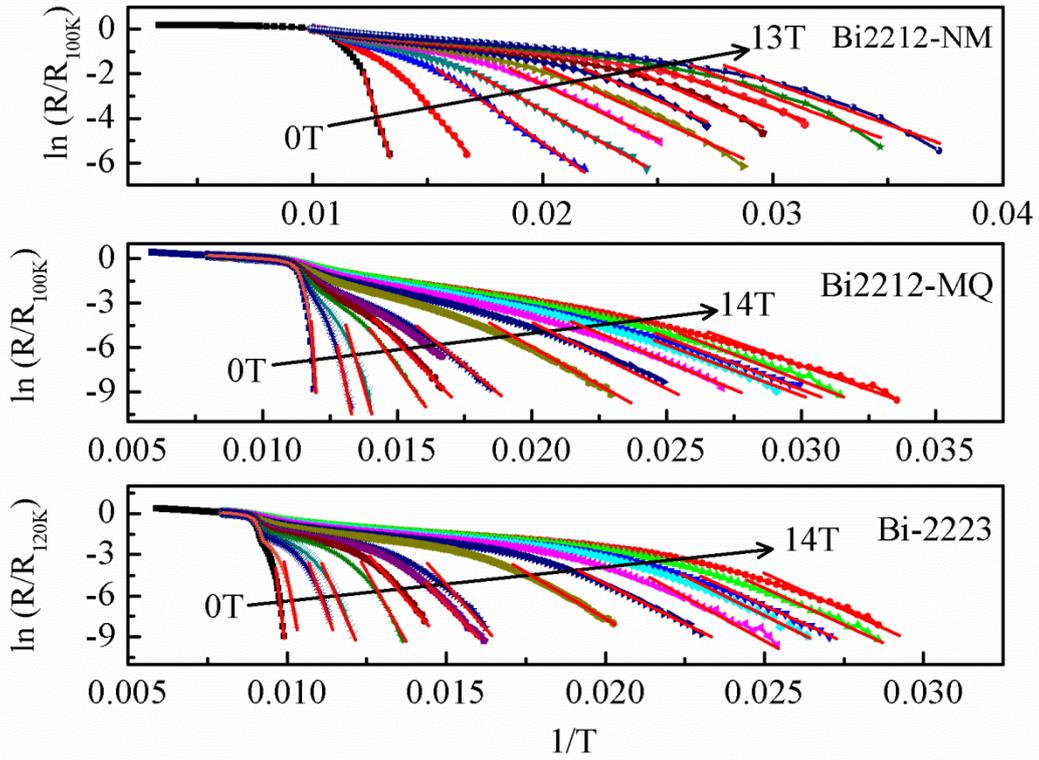

Figure 7

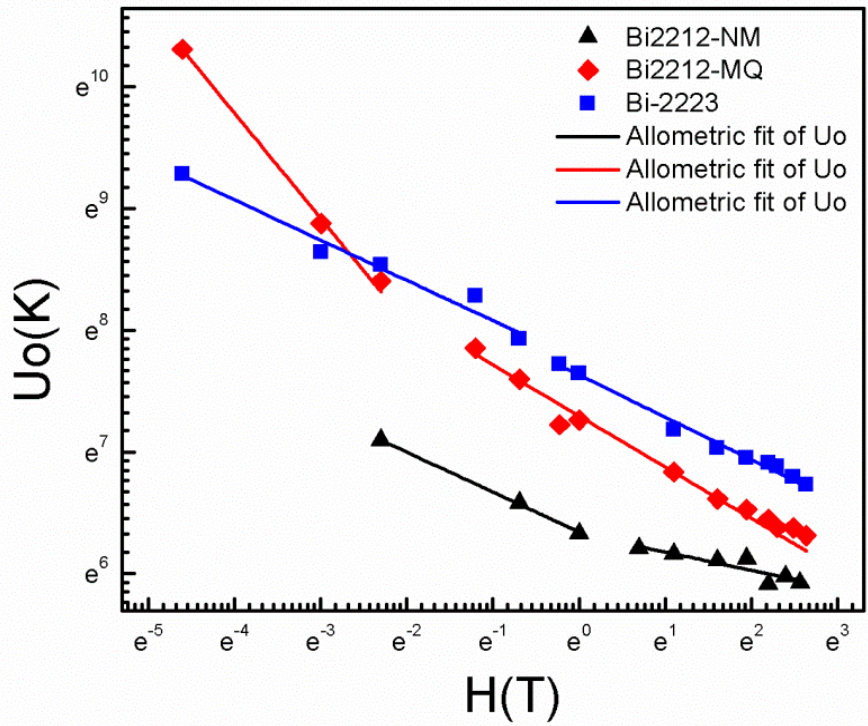

Figure 8

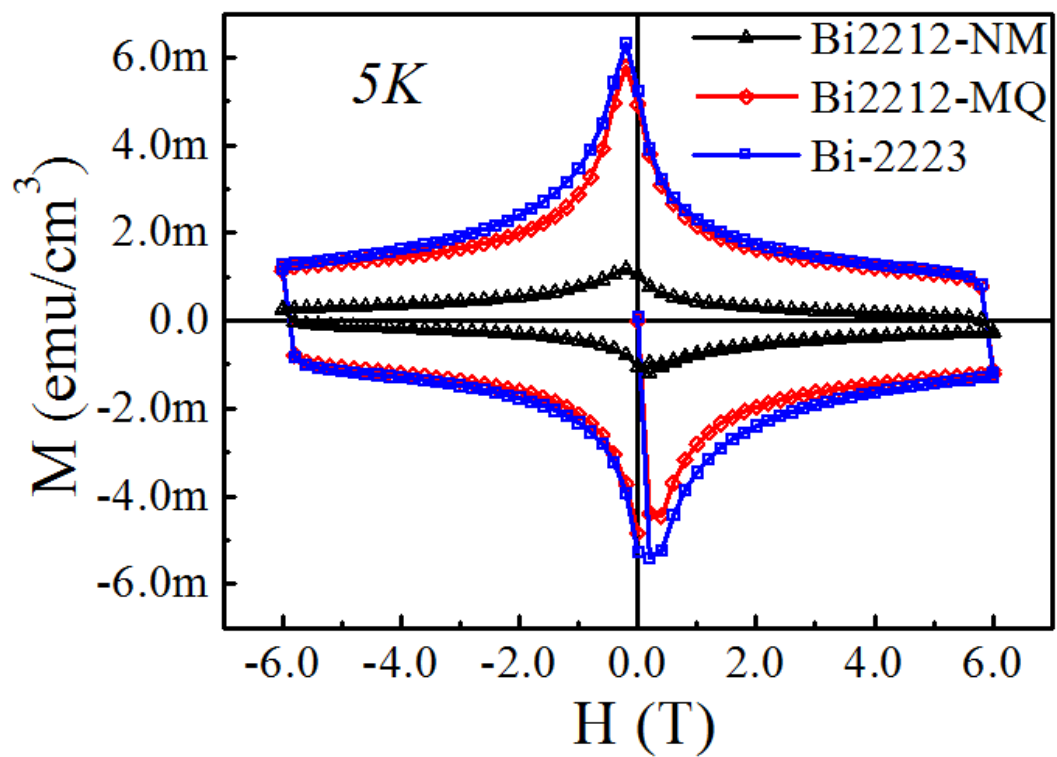

Figure 9

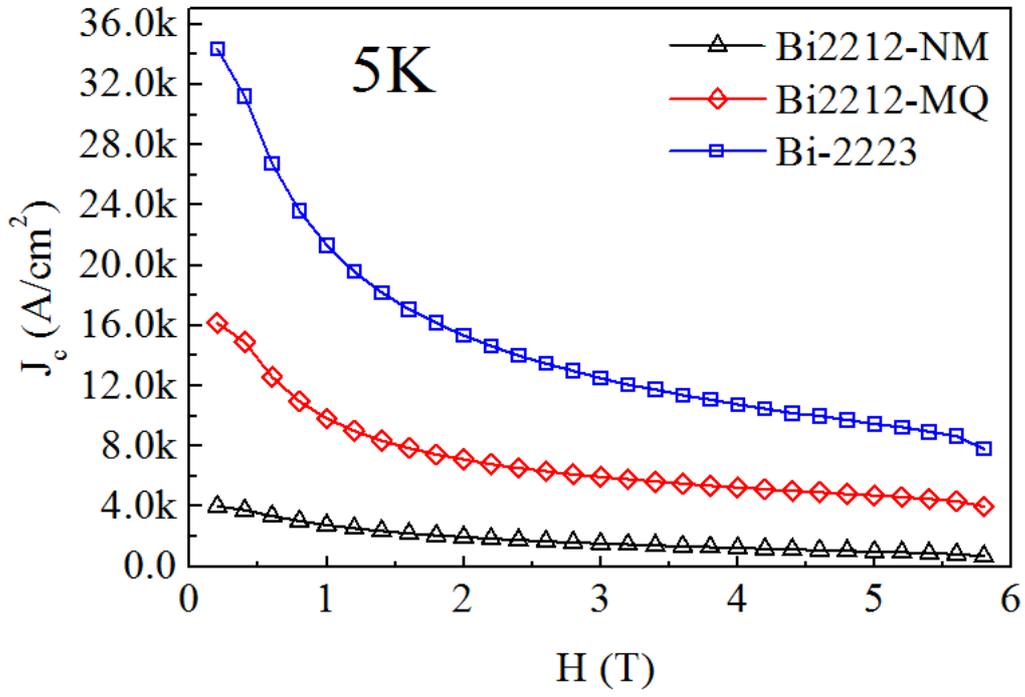

Figure 10

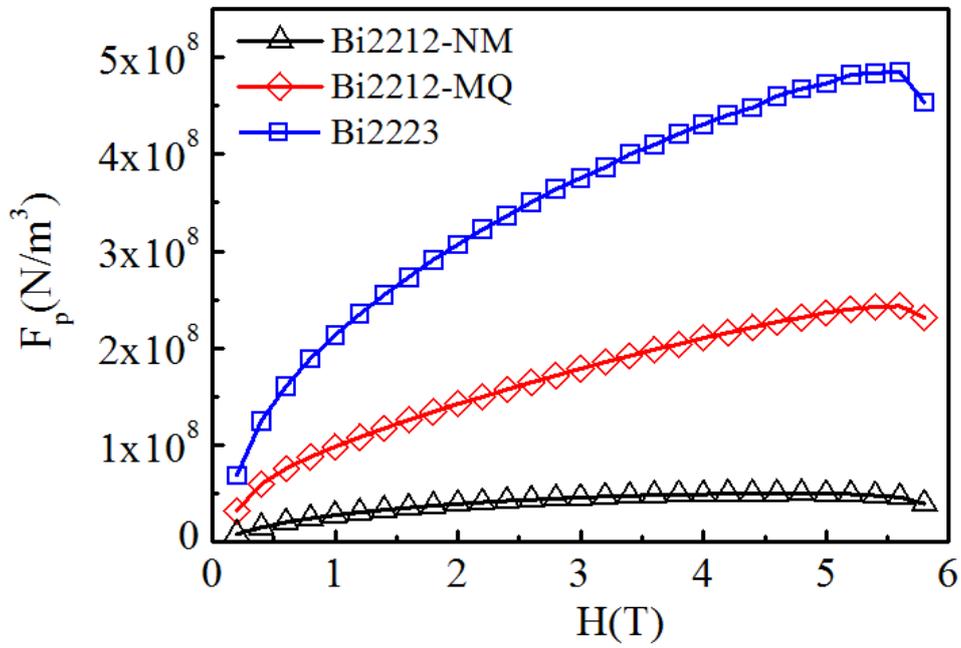

Figure 11 (a) and (b)

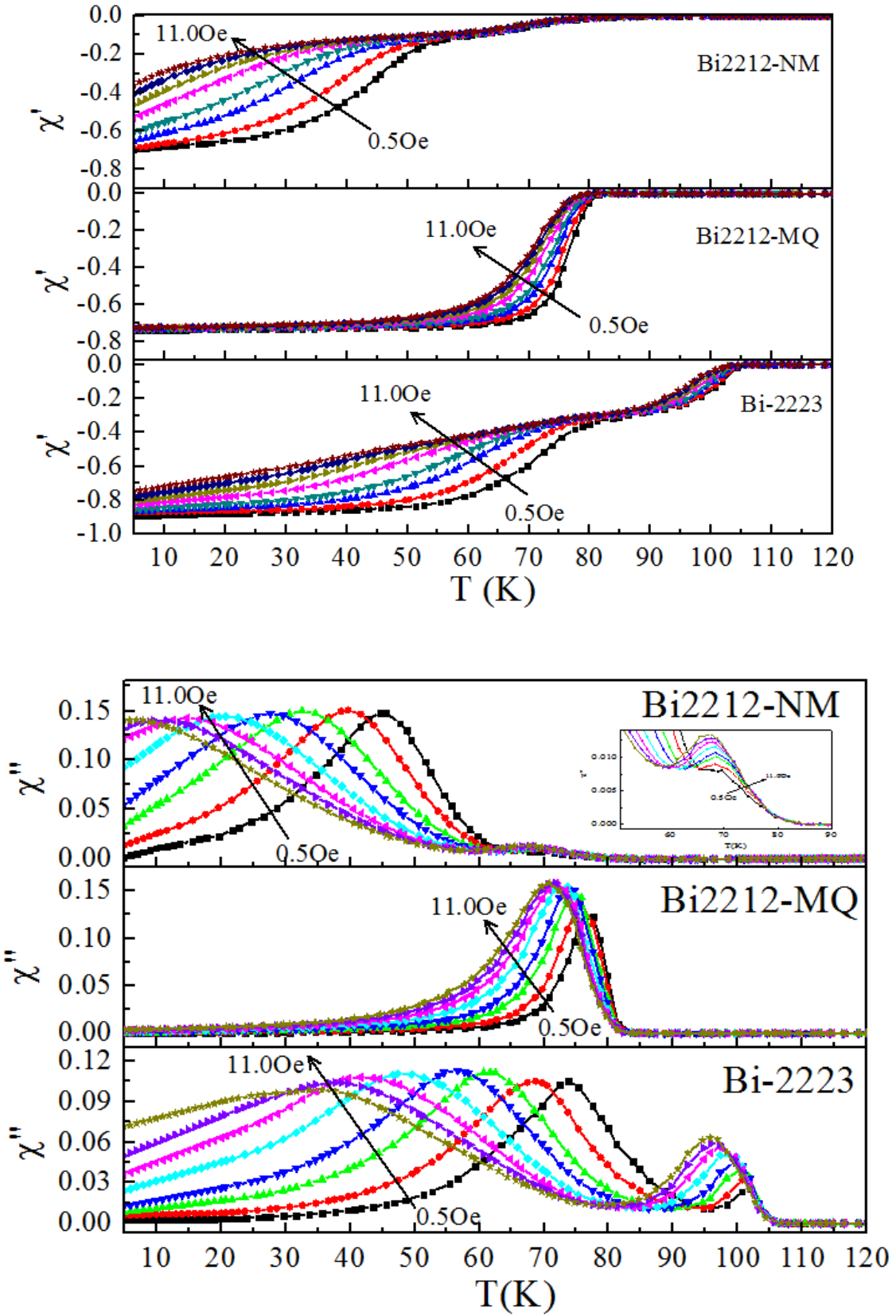

Figure 12 (a) to (c)

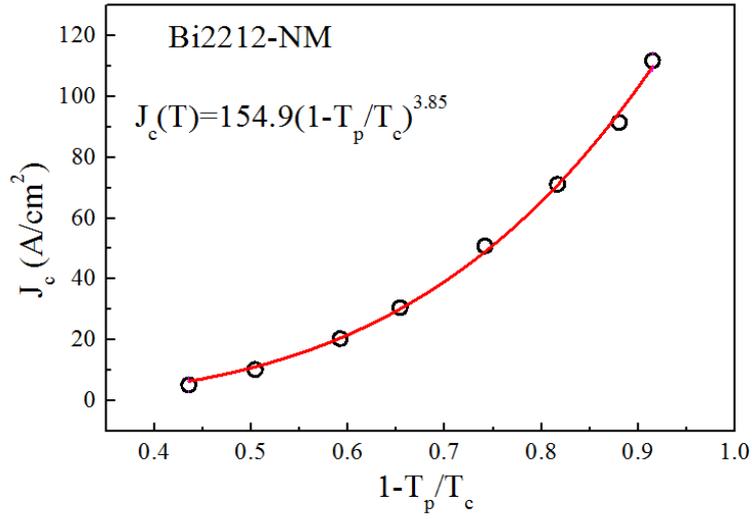

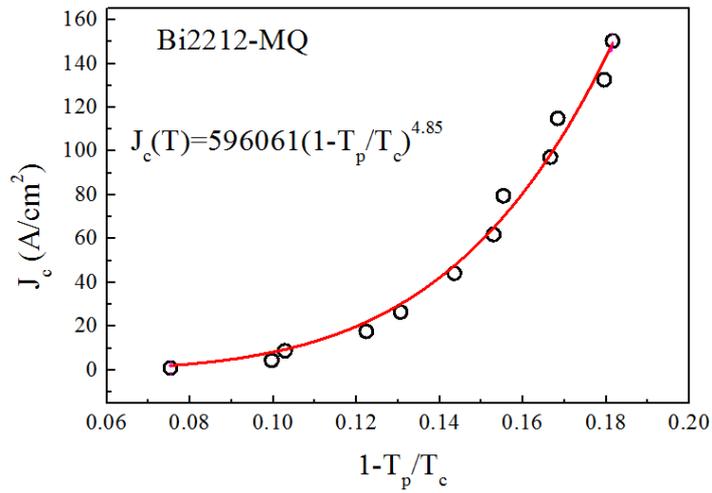

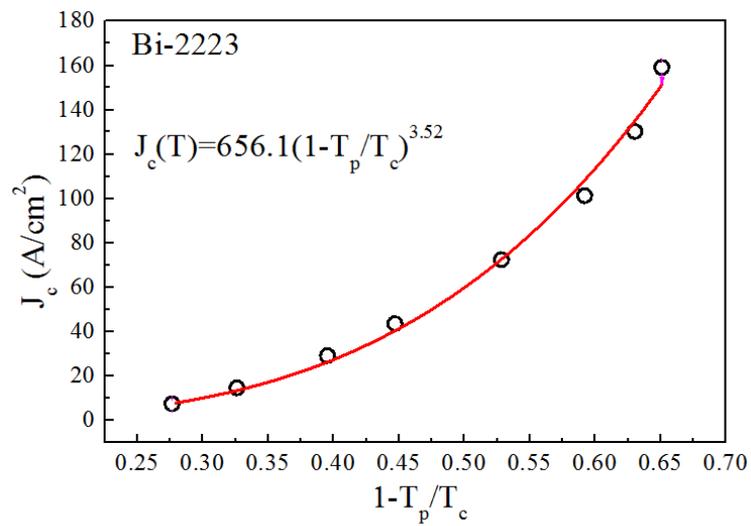

Figure 13 (a) and (b)

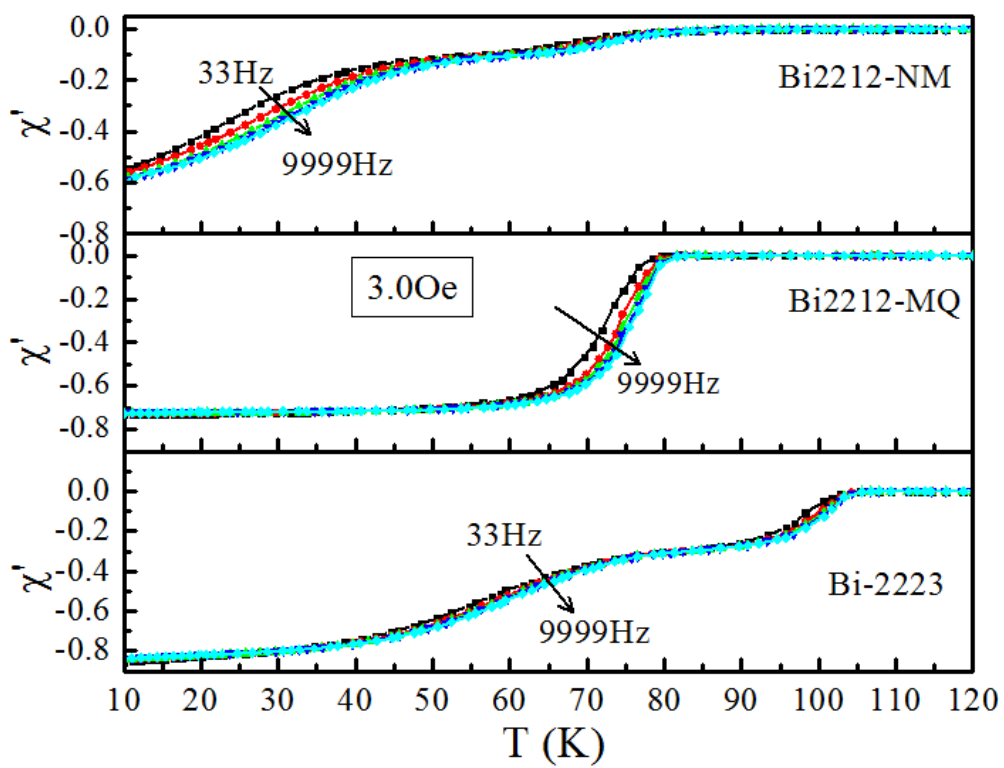

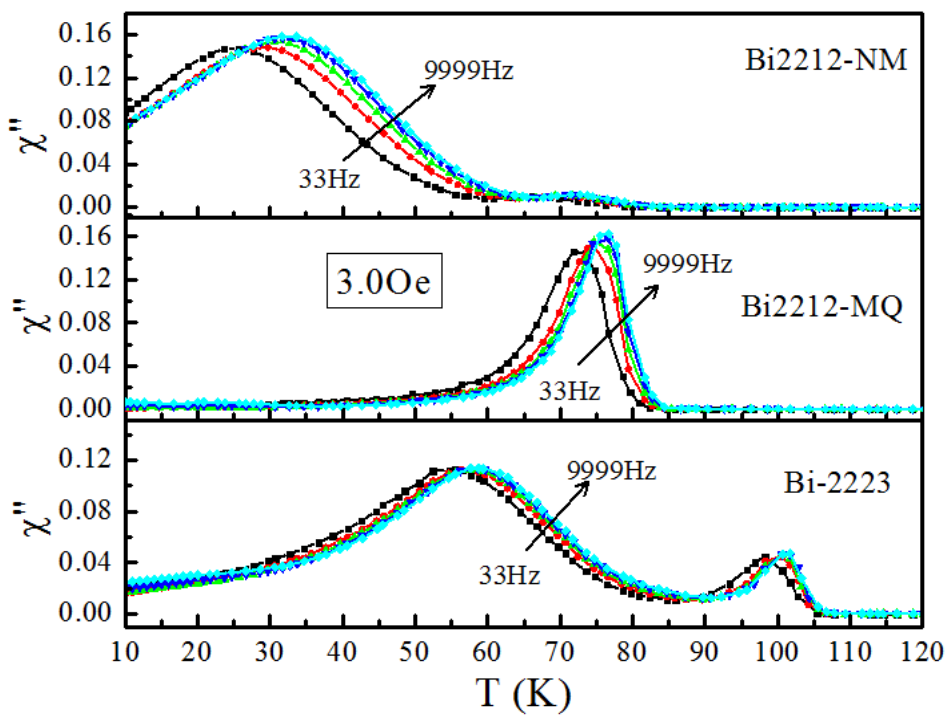

Figure 14 (a) to (c)

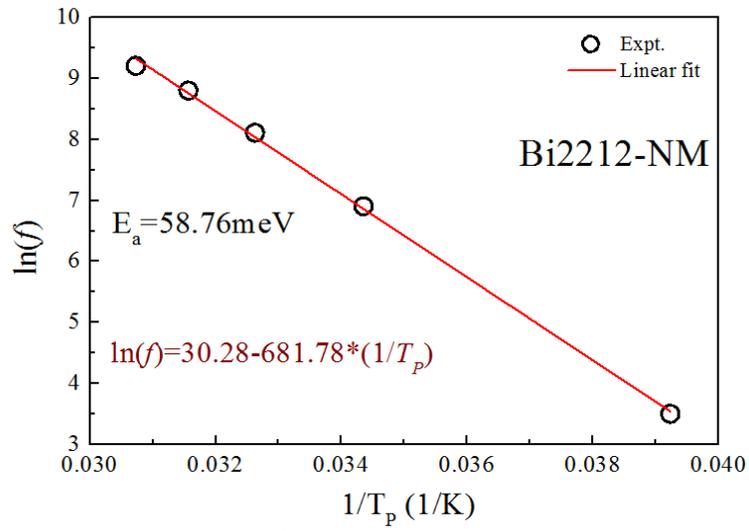

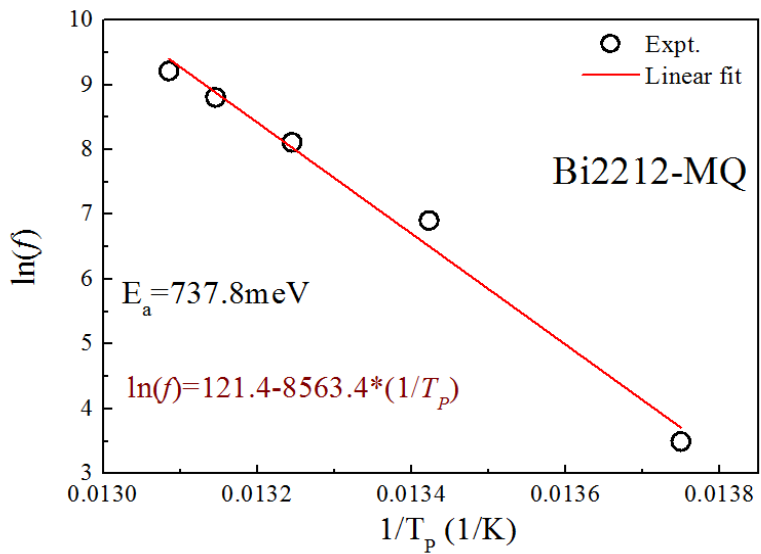

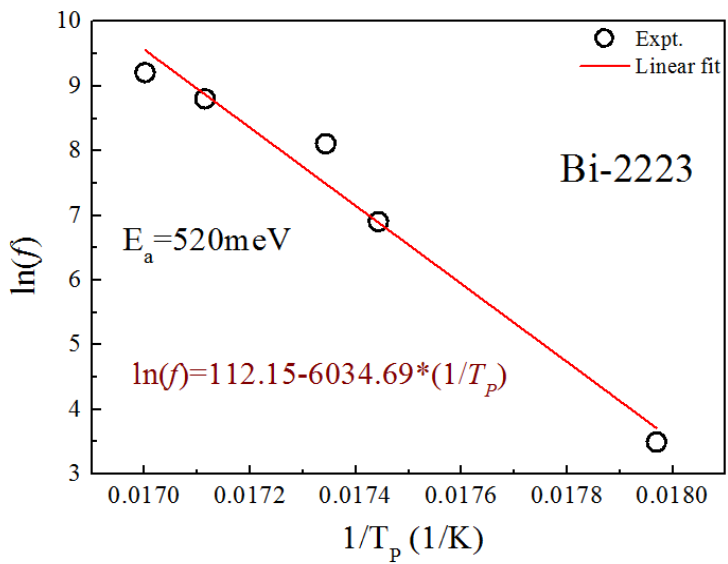